\documentclass[%
 aip,
 amsmath,amssymb,
 reprint,%
]{revtex4-1}

\usepackage{graphicx}
\usepackage{dcolumn}
\usepackage{bm}

\usepackage[utf8]{inputenc}
\usepackage[T1]{fontenc}
\usepackage{mathptmx}
\usepackage{etoolbox}
\usepackage{comment}
\usepackage{cases}
\usepackage[labelfont=bf,textfont=normalfont,singlelinecheck=off,justification=raggedright]{caption}
\usepackage{float}
\usepackage{extarrows}
\usepackage{subfig}
\makeatletter
\def\@email#1#2{%
 \endgroup
 \patchcmd{\titleblock@produce}
  {\frontmatter@RRAPformat}
  {\frontmatter@RRAPformat{\produce@RRAP{*#1\href{mailto:#2}{#2}}}\frontmatter@RRAPformat}
  {}{}
}%
\makeatother

\newtheorem{ansatz}{Ansatz}
\begin{document}

\preprint{AIP/123-QED}

\title{Opinion dynamics on biased dynamical networks: beyond rare opinion updating}
\author{Xunlong Wang}
 
\author{Bin Wu}
 \email{bin.wu@bupt.edu.cn}
\affiliation{ 
\begin{tabular}{c}
School of Science, Beijing University of Posts and Telecommunications, Beijing 100876, China
\end{tabular}
}%

\begin{abstract}
Opinion dynamics is of paramount importance as it provides insights into the complex dynamics of opinion propagation and social relationship adjustment. It is assumed in most of the previous works that social relationships evolve much faster than opinions. This is not always true in reality. We propose an analytical approximation to study this issue for arbitrary time scales between opinion adjustment and network evolution. 
   To this end, the coefficient of determination in statistics is introduced and a one-dimensional stable manifold is analytically found, i.e., the most likely trajectory.  With the aid of the stable manifold, we further obtain the fate of opinions and the consensus time, i.e., fixation probability and fixation time. 
    We find that for in-group bias, the more likely individuals are to adopt the popular opinion, the less likely the majority opinion takes over the population, i.e., conformity inhibits the domination of popular opinions. 
   This counter-intuitive result can be interpreted from a game perspective, in which in-group bias refers to a coordination game and rewiring probability refers to a rescaling of the selection intensity. 
   Our work proposes an efficient approximation method to foster the understanding of opinion dynamics in dynamical networks.
\end{abstract}

\maketitle

\begin{quotation}
Opinion formation on dynamical networks mirrors the interdependence of human behavior and the ever-changing social network,
which is of significant importance across various disciplines including computer science, physics and sociology.
Individuals can either adjust their social relationships or their opinions.
Herein, the frequency with which individuals adjust their opinions, measuring the strength of conformity,
is crucial for the opinion formation.
It is still unclear whether such frequency would promote consensus formation or not.
We establish a framework of opinion formation on dynamical networks to show that conformity inhibits the majority opinion from taking over the population for in-group bias. We also provide an explanation of the counter-intuitive results from a game perspective, although we do not assume any game in prior.
Our work sheds light of opinion formation with the aid of game theory,
which bridges the gap between the two fields.

\end{quotation}

\section{\label{sec:introduction} Introduction}
Opinion dynamics is an interdisciplinary field that offers a multifaceted understanding of how opinions and beliefs shape our world \cite{bonabeau1999swarm,andris2015rise,vinokur1978depolarization,ojer2023modeling}. 
It has received wide and ample attention for half a century \cite{perc2023signal,weidlich1971statistical}.
Its significance extends to physics, computer science, economics and social sciences \cite{perc2022social}.
Statistical physics provides various efficient toolboxes to analyze and predict opinion formation. 

Opinion dynamics on social networks has been intensively studied \cite{wu2020CCC,degroot1974reaching,zschaler2012early,baumann2020modeling,du2022voter,wang2020public,liggett1999voter}. 
Besides, individuals' social relationships aren't set in stone. 
Individuals can voluntarily stay away from those who do not adopt their opinions corresponding to in-group bias \cite{holme2006nonequilibrium,durrett2012graph,fu2012evolution}. And individuals can be prone to stay away from those with similar opinions corresponding to out-group bias \cite{kimura2008coevolutionary}.
Thus social relationships are under social biases.
Both opinion propagation and social relationship adjustment thus need to be considered simultaneously, which refers to dynamical networks. 
In fact, dynamical networks have been studied for about two decades in various fields, such as opinion dynamics \cite{gross2008adaptive,wu2020CCC,du2022voter,AT2006activelink,fu2008coevolutionary} and evolutionary game theory \cite{perc2010coevolutionary,du2022evolutionary,du2023coevolutionary,du2024asymmetric}. And opinion dynamics has provided insights for opinion formation in the real world ranging from consensus to polarization \cite{holme2006nonequilibrium,kimura2008coevolutionary,baumann2020modeling,wang2020public,durrett2012graph,fu2008coevolutionary,fu2018opinion}.

It is typically assumed in dynamical network models \cite{wu2020CCC,shan2022social,AT2006activelink,wu2010evolution,krosnick1988attitude,wang2023opinions,liu2023emergence,wu2019evolution,wu2011evolutionary,wu2016evolving,baumann2020modeling,wei2019vaccination} that social relationships evolve much faster than opinion updating. 
This assumption has facilitated the analysis.
The assumption has two implications. 
On one hand, it implies that individuals stubbornly stick to their opinions. 
In reality, they don't adhere to their own opinions. For example, personal opinions are constantly shaped by diverse news in the face of closely contested matches between evenly-matched teams.
On the other hand, it implies that individuals prefer to adjust their social relationships rather than to change their opinions. This is not true. For example, when individuals are shifting about what to eat later, they are prone to be influenced by the opinions of their peers, but it is less likely to break off the relationships due to the difference of opinions.
Therefore, it is worthwhile to investigate a dynamical network model in which the evolution of opinions and social relationships occur at a similar time scale.


We concentrate on the edge dynamics with probabilistic breakup and random rewiring \cite{durrett2012graph,wu2010evolution,wang2023opinions,shan2022social,wu2020CCC,du2022voter}.
In fact, a social connection between any two individuals is not perfectly strong. So we suppose that the breakup of any social relationship is likely to happen instead of only the breakup of heterogeneous edges (i.e., two ends of the edge hold different opinions) \cite{durrett2012graph,holme2006nonequilibrium,kimura2008coevolutionary,rogers2013consensus,simplex2019}.
 We are addressing the following questions: i) what is the most likely trajectory to consensus? ii) what is the likelihood that a given opinion takes over the population? iii) how long does it take for the population to reach consensus?
 
\section{Model}

We propose a co-evolutionary network model incorporating opinion dynamics on the network and dynamics of the network structure.

    Let us consider a network which consists of $N$ nodes representing individuals
    and $L$ edges representing social ties. Each node holds either opinion $A$ or opinion $B$. 
    The types of edges are of three types, which are $AA$, $AB$, and $BB$. In addition, we denote the average degree of the network by $\bar{k}=2L/N$. In this paper, we only study the connected network.
    Otherwise, the network has to be disconnected, and opinion propagation in such disconnected networks is of little significance.

    In each time step, the evolution of network structure (Process $1$) occurs with probability $1-\phi$, and the evolution of opinions (Process $2$) occurs with probability $\phi$ where $0<\phi<1$.
    Next, we describe the Process $1$ and Process $2$ separately.

    Let us consider the dynamics of the network structure (Process $1$). An edge is randomly selected.
    If the two nodes connected by this edge hold different opinions (i.e., the type of the edge is $AB$), the edge breaks with probability $k_d$. 
    If the two nodes connected by this edge hold the same opinions (i.e., the type of the edge is $AA$ or $BB$), the edge breaks with probability $k_s$. 
    If the selected edge doesn't break, nothing happens, and 
    Process $1$ finishes. 
    Otherwise, one of the two ends of the broken edge is selected randomly and rewires to another node at random that is not in its neighborhood to avoid multiple edges. In other words, one edge breaks off and another emerges.

    Let us consider the opinion dynamics on the network  (Process $2$).
    We adopt the voter model: a node $i$ is selected randomly. If node $i$ has no neighbors, nothing happens, and Process $2$ finishes. Otherwise, a node $j$ is selected randomly in node $i's$ neighborhood. Node $i$ adopts the opinion of node $j$. Process $2$ is equivalent to a death-birth process in the structured network under neutral selection \cite{altrock2009fixation}. 

    Isolated network structures
    appear from time to time during the rewiring
    process.
    They are sure to disappear since isolated sub-networks will be connected via the rewire-to-random property of Process $1$. Thus, opinions spread to every corner of the network sooner or later. 
    In other words, consensus (i.e., All $A$ or All $B$) is the only stationary state.

    We are interested in how the entire network reaches consensus. In the next section, we will show i) the most likely trajectory via which the system reaches consensus; ii) the probability of consensus of a given opinion; and iii) the time taken by the system to reach consensus.

\section{pair-approximations \& mean-field equations\label{Chapter: Method}}

$N$, $L$, $\bar{k}$ are constants because neither Process $1$ nor Process $2$ changes the total number of nodes and the total number of edges. Nevertheless, the proportion of various opinions and the proportion of edges of various types are changing.
    
    We denote the number of nodes holding opinion $X$ by $[X]$ and the number of $X-Y$ edges by $[XY]$, where $X, Y \in \{A, B\}$ and $[XY]=[YX]$. We are able to capture the system by the following five variables $[AA]$, $[BB]$, $[AB]$, $[A]$, and $[B]$. Note that we only need to study three independent variables $[AA]$, $[BB]$ and $[A]$ since $[AA]+[BB]+[AB]\equiv L$ and $[A]+[B]\equiv N$.
    Triplets (i.e. $X-Y-Z$, where the central $Y$ has $X-Y$ and $Y-Z$ edges) are introduced. Denote the number of $X-Y-Z$ triplets by $[XYZ]$. Due to the symmetry of triplets, we have that $[XYZ]=[ZYX]$.

    For $[AA]$, both Process $1$ and Process $2$ can change it. Let us first consider how Process $1$ influences $[AA]$. If an $A-B$ edge is selected and breaks off, and if the node holding opinion $A$ at the ends of the broken edge is selected and rewires to a node holding opinion $A$ in the network, then $[AA]$ increases by one. If an $A-A$ edge is selected and breaks off, the selected node of the broken edge has to be $A$, and the selected node rewires to another node in the network that holds opinion $B$, then $[AA]$ decreases by one. Let us secondly consider how Process $2$ influences $[AA]$. If a node holding opinion $B$ is selected, and if a node in its neighborhood who holds opinion $A$ is selected too, and if the former node adopts the opinion of the latter one, then $[AA]$ increases by one. If a node with opinion $B$, which has two neighbors holding opinion $A$, is selected (i.e.,  the node in the center of a triplet $A-B-A$ is selected), and if the center node adopts the opinion of any neighbor holding opinion $A$, then $[AA]$ increases by two. If a node with opinion $A$, which has one neighbor with opinion $A$ and another neighbor with opinion $B$, is selected (i.e., the node in the center of a triplet $A-A-B$ is selected), and if the center node adopts the opinion of the neighbor holding opinion $B$, then $[AA]$ decreases by one. Therefore, we obtain the mean-field equation of $[AA]$,
    \begin{align}
        \frac{d[AA]}{dt}&\!=\!\left\{\! \frac{[B]}{N}R(A|B)\!+\! 2\frac{[B]}{N}R(ABA|B)\!-\!\frac{[A]}{N}R(AAB|A)\!\right\}\nonumber\\
        &\!\times\!\phi\!+\!\left\{\! \frac{[AB]}{L}k_d\frac{1}{2}\frac{[A]}{N}\!-\!\frac{[AA]}{L} k_s \frac{[B]}{N}\!\right\}\!\times\!(1\!-\!\phi).\label{eq:1}
    \end{align}

    \begin{figure*}[ht]
        \centering
        \subfloat{\includegraphics[width=.5\textwidth]{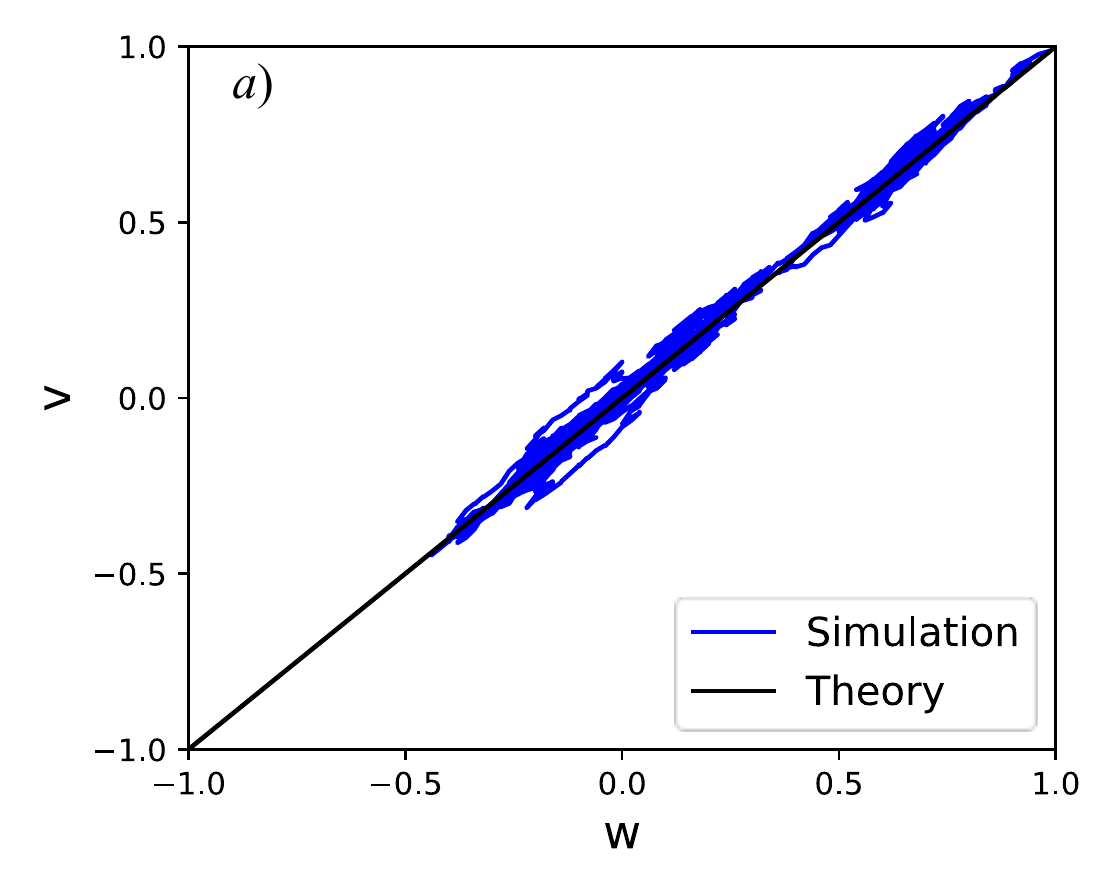}}
        \subfloat{\includegraphics[width=.5\textwidth]{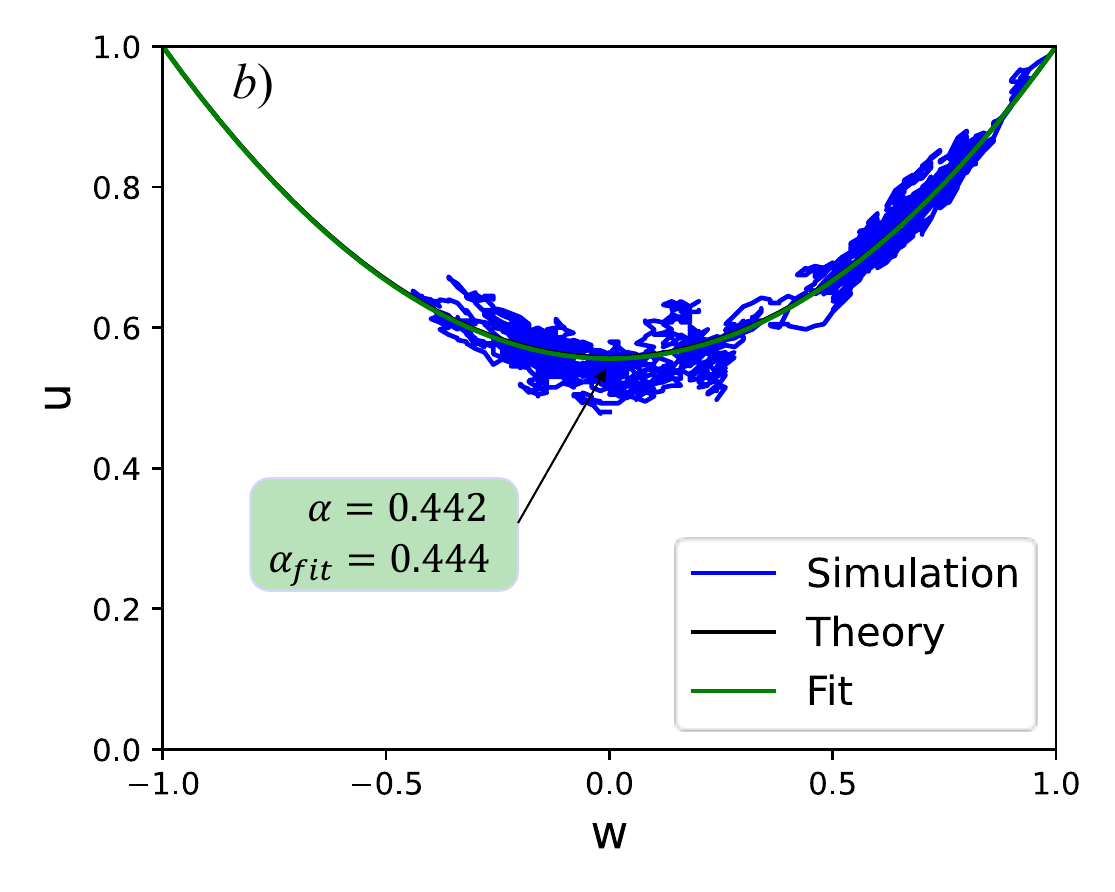}}
        \caption{\textbf{The system converges to the stable manifold and takes a random walk on it.} (a) $v$ and (b) $u$ as functions of $w$, i.e., the stable manifold Eqs. (\ref{eq:7}-\ref{eq:8}). It is a simulation for the case where $k_d=k_s=0.6$ and $\phi=0.5$. In all the panels, the population size is $N = 100$ and $L=400$. The population starts
        with half opinion $A$ and half opinion $B$. The initial network structure is generated randomly and the initial edges are placed randomly. The green line is the best fit of the path to the parameterized parabola $u=1-\alpha_{fit}(1-w^2)$. In order to avoid the effect of the pre-convergence trajectory, we fit with the last ninety percent of the trajectory. The difference between the theoretical $\alpha$ and the fitting $\alpha_{fit}$ is of order $10^{-3}$, which is very small.}
        \label{fig: uv1}
    \end{figure*}

    Given a node with opinion $X$ is selected, $R(Y|X)$ denotes the rate at which the node with opinion $X$ finds a neighbor with opinion $Y$;
    $R(YXZ|X)$ denotes the rate at which the node with opinion $X$ randomly chooses two neighbors, one of which is with opinion $Y$ and the other of which is with opinion $Z$ (i.e., the selected node is in the center of a triplet $Y-X-Z$).   

    Similarly, we write the mean-field equations of $[BB]$ and $[A]$ based on the co-evolutionary process,
    \begin{align}
        \frac{d[BB]}{dt}&\!=\!\left\{\! \frac{[A]}{N}R(B|A)\!+\! 2\frac{[A]}{N}R(BAB|A)\!-\!\frac{[B]}{N}R(ABB|A)\!\right\}\nonumber\\
        &\!\times\!\phi\!+\!\left\{\!\frac{[AB]}{L}k_d\frac{1}{2}\frac{[B]}{N}\!-\!\frac{[BB]}{L}k_s \frac{[A]}{N}\!\right\}\!\times\!(1\!-\!\phi),\label{eq:2}\\
        \frac{d[A]}{dt}&\!=\!\phi\!\times\!\left\{\!\frac{[B]}{N}\cdot R(A|B)\!-\!\frac{[A]}{N}\cdot R(B|A)\!\right\}.\label{eq:3}
    \end{align}

    Eqs. (\ref{eq:1}-\ref{eq:3}) are closed.

    We use the pair-approximation. In other words, we assume that $R(Y|X)=[XY]/(\sum_{W\neq X}[XW]+2[XX])$ and 
    $R(YXZ|X)=[YXZ]/(\sum_{W\neq X}[XW]+2[XX])$ where $[XYX]=[XY][XY]/(2[Y])$ and $[XXY]=2[XX][XY]/[X]$ \cite{kimura2008coevolutionary}.

    In addition, we adopt a coordination transformation. Setting $u=([AA]+[BB])/L$, $v=([AA]-[BB])/L$, and $w=([A]-[B])/N$, Eqs. (\ref{eq:1}-\ref{eq:3}) become
    \begin{numcases}{}
    \frac{du}{dt}=\phi\frac{2}{N\bar{k}}\cdot\frac{1-u}{1-v^2}\left[1-vw+\bar{k}(1-2u+v^2)\right]\nonumber\\
    \quad\quad\ +\frac{1-\phi}{N\bar{k}}\left[k_d(1-u)-k_s(u-vw)\right],\label{eq:4}\\
    \frac{dv}{dt}=\phi\frac{2}{N\bar{k}}\cdot\frac{1-u}{1-v^2}(v-w)\nonumber\\
    \quad\quad\ +\frac{1-\phi}{N\bar{k}}\left[k_dw(1-u)-k_s(v-uw)\right],\label{eq:5}\\
    \frac{dw}{dt}=\phi\frac{2}{N}\cdot\frac{1-u}{1-v^2}\cdot(v-w)\label{eq:6}
    \end{numcases}
    where $0\le u\le 1$, $-1\le v\le 1$ and $-1\le w\le 1$. 
    
    The variable $u$ represents the proportion of homogeneous edges (i.e., both ends of the edge hold the same opinion).
    If $u>1/2$, then more than half of the social ties are between those with the same opinion. The variable $w$ represents the popularity of opinion $A$. When $w$ is positive, opinion $A$ is more popular than opinion $B$. When $w$ is equal to $1$, every node in the population holds opinion $A$ (i.e., All $A$). 
    Every node holds opinion $B$ (i.e., All $B$) when $w$ is equal to $-1$.

\section{Fixation via the most likely trajectory}
\subsection{Benchmark case, $k_d=k_s$}
We consider a benchmark case, where $k_s=k_d=k_0$. We find that a one-dimensional stable manifold emerges in Eqs. (\ref{eq:4}-\ref{eq:6}) (See Appendix \ref{Appendix:manifold} for the derivation),
    \begin{numcases}{}
        u = 1-\alpha(1-w^2),\label{eq:7}\\
        v = w,\label{eq:8}
    \end{numcases}
    where $\alpha=\frac{\phi(\bar{k}\!-\!1)\!-\!(1\!-\!\phi)k_0\!+\!\sqrt{k_0^2(1\!-\!\phi)^2\!+\!2(1\!+\!\bar{k})k_0\phi(1\!-\!\phi)\!+\!\phi^2(\bar{k}\!-\!1)^2}}{4\phi\bar{k}}$  has nothing to do with the three variables $(u,v,w)$, and only depends on the average degree $\bar{k}$, the breaking probabilities $k_0$, and the probability that Process $2$ occurs $\phi$.

    The stability of the manifold implies that the system converges to the quadratic curve depicted by Eqs. (\ref{eq:7}-\ref{eq:8}), no matter what the initial network configuration and the popularity of opinions are (as shown in Fig.~\ref{fig: uv1}).

    The stable manifold further facilitates us to capture the co-evolutionary process as a one-dimensional Markov chain with state variable $w$, or equivalently the number of nodes with opinion $A$ (i.e., $[A]$).    
    Interestingly, the change of $[A]$ is influenced only by Process $2$, rather than Process $1$.    
    The state space of the one-dimensional Markov chain is thus $\Omega=\{0,1,2,\cdots, N\}$. Each state represents the number of nodes with opinion $A$.    
    Herein both $0$ (All $B$) and $N$ (All $A$) are the absorbing states of the Markov chain.
    The transition probabilities are given by
    \begin{align}
        P^{+}_j=P^{-}_j=\frac{\alpha\phi}{2}[1-(\frac{2j}{N}-1)^2],\label{eq:9}
    \end{align}
    where $P^{+}_j$ ($P^{-}_j$) denotes the probability that the Markov chain increases (decreases) by one from state $j$. (See Appendix \ref{Appendix: fixation} for the derivation)

    \begin{figure}[tbp]
        \centering
        \includegraphics[width=\columnwidth]{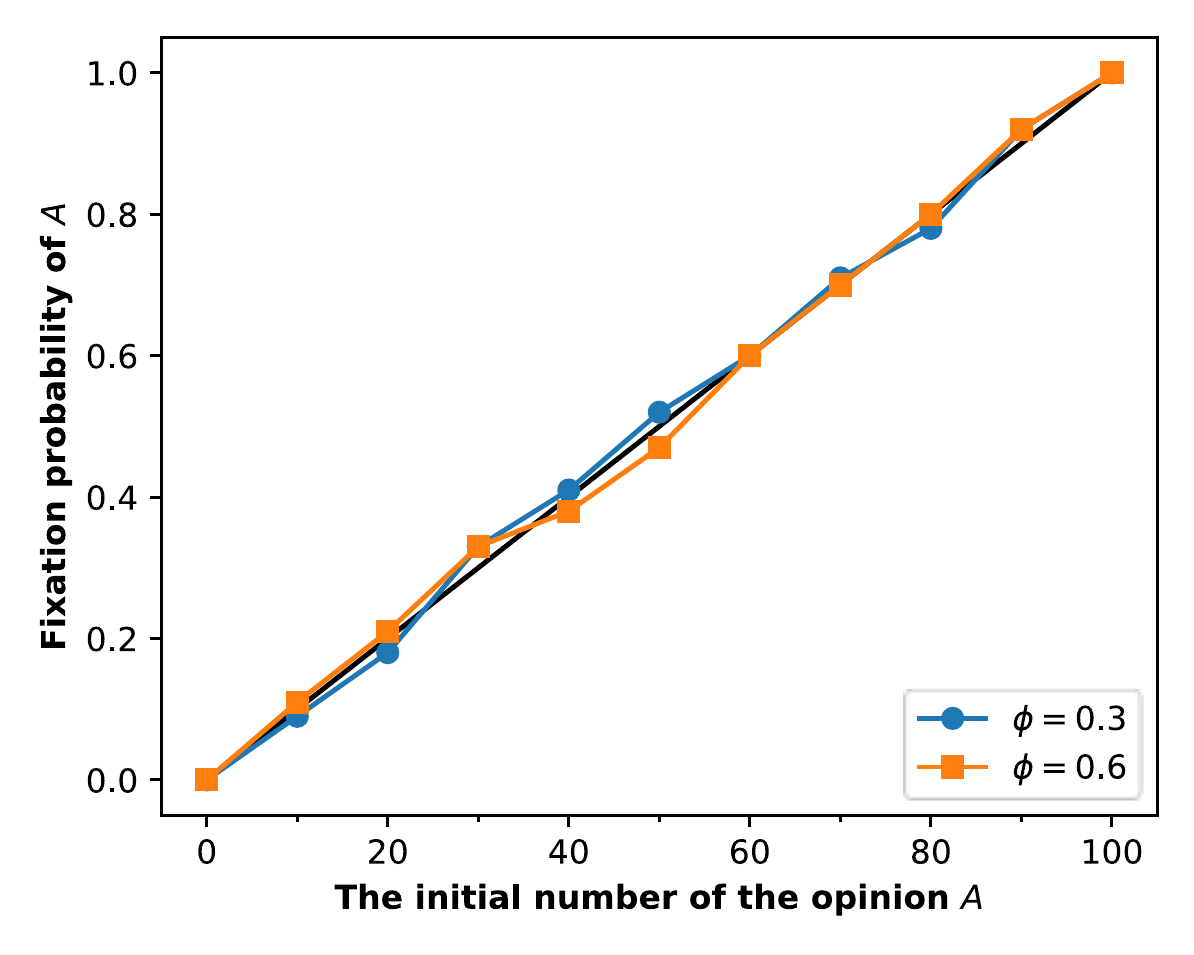}
        \caption{\textbf{The fixation probability is independent of $\phi$ for $k_d=k_s$.} It's found that the fixation probability varies linearly with the initial number of the opinion $A$ in the benchmark case, which is a fingerprint of the random walk on the stable manifold. The dots (square or circle) represent the simulation results and the simulation data points are calculated by averaging over $100$ independent runs. Here $k_d=k_s=0.6$. Different values of $\phi$ have no effect on the fixation probability.}
        \label{fig: bp}
    \end{figure}

    To capture the fate of the system, we are interested in the likelihood that $j$ opinion $A$ nodes take over the system. 
    Fixation probability (Consensus probability) $\Phi_j$ is the probability that the Markov chain starts from state $j$ and is eventually absorbed by state $N$. The formula for the fixation probability is given by $\Phi_j=\frac{j}{N}$ (See Appendix \ref{Appendix: fixation} for the derivation). The fixation probability thus has nothing to do with the likelihood of Process $2$ (i.e. $\phi$), and the average degree (i.e., $\bar{k}$) provided that the two breaking probabilities are the same (as shown in Fig.~\ref{fig: bp}).

    The result is equivalent to the fixation probability in the fixed regular graph and the complete graph\cite{maciejewski2014reproductive}. The fixation probability of our co-evolutionary model is independent of $\phi$, which implies speed of change in network structure has nothing to do with the likelihood of which opinion takes over the system. Intuitively, if $\phi\to 0^+$, the fixation probability on the evolving network is equivalent to that of the Moran process on the complete network with payoff matrix $\begin{bmatrix}
        \frac{1}{k_s} & \frac{1}{k_d}\\
        \frac{1}{k_d} & \frac{1}{k_s}
    \end{bmatrix}$\cite{wu2020CCC}. If $k_s=k_d=k_0$, then the game refers to neutral selection. Our results indicate the robustness of the result beyond $\phi\to 0^+$.

    \begin{figure}[tp]
        \centering
        \includegraphics[width=\columnwidth]{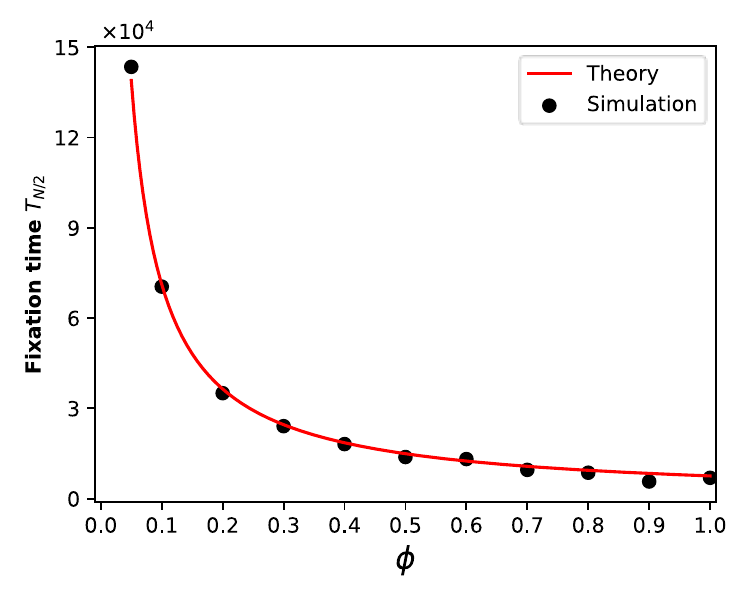}
        \caption{\textbf{Simulation results coincide perfectly with the theoretical results.} For every realization, the population starts
        with half opinion $A$ and half opinion $B$, and the initial edges are placed randomly. The simulation data points are calculated by averaging over $100$ independent runs. Here $k_d=k_s=0.6$. }
        \label{fig: bt}
    \end{figure}

    Fixation time (Consensus time) $T_j$ denotes the total time that the Markov chain starts from state $j$ and is eventually absorbed by state $0$ or state $N$. 
    The fixation time $T_j$ in general is approximated by $\left\{\!(N^2\!-\!N)\ln(N\!-\!1)\!-\!N[(N\!-\!j)\ln(N\!-\!j)\!+\!j\ln j]\!\right\}/(2\alpha\phi)$.
    (The  derivations of the invasion time $T_1$ and the fixation time $T_j$ are provided in the Appendix \ref{Appendix: fixation})

    Although all derivations are based on the case for $0<\phi<1$, the results are robust and almost valid for the case for $\phi=1$. 
    When $\phi$ is equal to $1$ in the benchmark case, our co-evolutionary model is equivalent to the Moran process under neutral selection on the static graph. 
    In this case, the theoretical solution of fixation time agrees with the outcomes for the regular graphs \cite{sui2015speed} and the complete graphs \cite{altrock2009fixation}.
    $T_j$ decreases with increasing $\phi$ (as shown in Fig.~\ref{fig: bt}),  which is intuitive because the speedup of opinion propagation contributes to consensus formation. However, it is counterintuitive that $T_j\phi$ increases with $\phi$. It implies that the adjustment of the network structure leads to consensus with minor opinion propagation.

    \begin{figure*}[t]
        \centering
        \subfloat{\includegraphics[width=.5\textwidth]{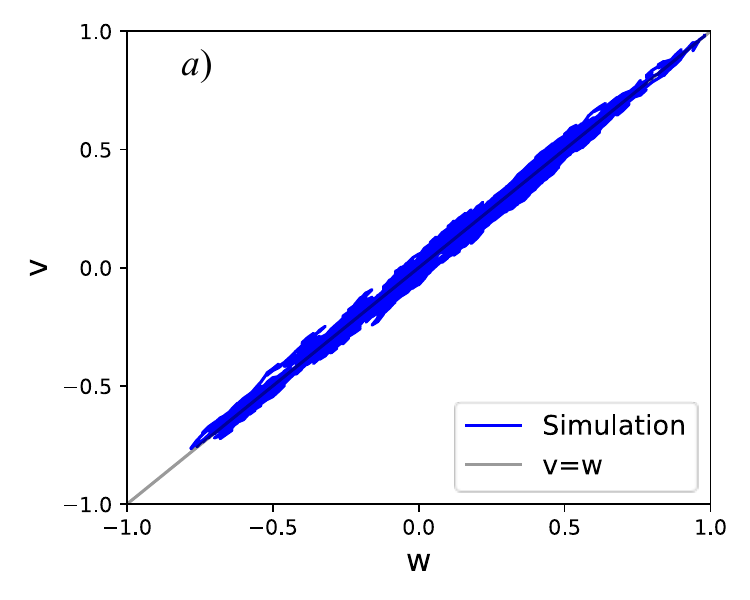}}
        \subfloat{\includegraphics[width=.5\textwidth]{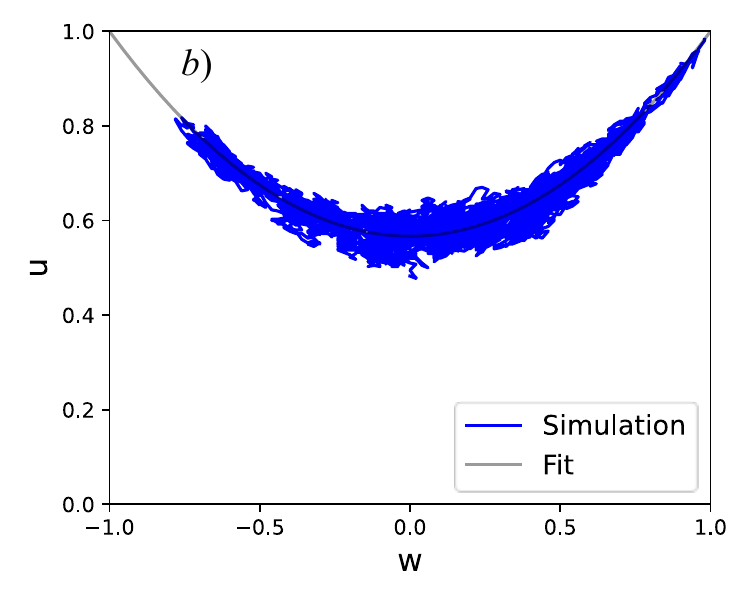}}
        \caption{\textbf{The manifold still exists and $v=w$ is still a good approximation of the most likely trajectory even if $k_d\neq k_s$.} It is a simulation for the case where $k_d=0.6$, $k_s=0.3$, and $\phi=0.6$. The dots represent the simulation results. (a) The line represents $v=w$. (b) The line is the best fit of the path to the parameterized parabola $u=1-\alpha_{fit} (1-w^2)$.}
        \label{fig: uv2}
    \end{figure*}

\subsection{General case, $k_d\neq k_s$}

In this section, we study the general case ($k_d\neq k_s$) where $k_d>k_s$ ($k_d<k_s$) corresponds to in-group (out-group) bias.

\subsubsection{Stable central manifold}

For $k_d\neq k_s$, the dynamical analysis of Eqs (\ref{eq:4}-\ref{eq:6}) leads to two fixed points $(u,v,w)=(1,\pm 1,\pm 1)$.
It implies that two consensus states (i.e., All $A$ and All $B$) are the stationary regimes.
On the other hand, no stable manifold is present based on Eqs. (\ref{eq:4}-\ref{eq:6}).
In fact, if $k_d\neq k_s$, then the $v=w$  doesn't hold due to Process $1$. 
However, it is shown by simulation (Fig.~\ref{fig: uv2}) that, there is still a stable manifold.
And  $v=w$ is still a good approximation of the stable manifold. In order to quantify the validity of the manifold $v = w$, we introduce the coefficient of determination \cite{dunn2018generalized,mccullagh1989generalized} $R^2=1-\sum(y_i-\hat{y})^2/\sum(y_i-\bar{y})^2$, which assesses the goodness of fit of a regression model.

We propose that $v=w$ holds well when $R^2$ is sufficiently large.
Let's denote the threshold as $R^*$.
If $R^2>R^*$, then the approximation $v=w$ is acceptable. 
Here $R^*$ is dependent on the population size $N$ and the average degree of the network $\bar{k}$,
and it has nothing to do with the breaking probabilities.

\begin{ansatz}
    If $R^2$ is sufficiently large, i.e., $R^2>R^*$, then $v=w$ is valid.\label{Ansatz1}
\end{ansatz}

\begin{figure}[htbp]
    \centering
    \includegraphics[width=\columnwidth]{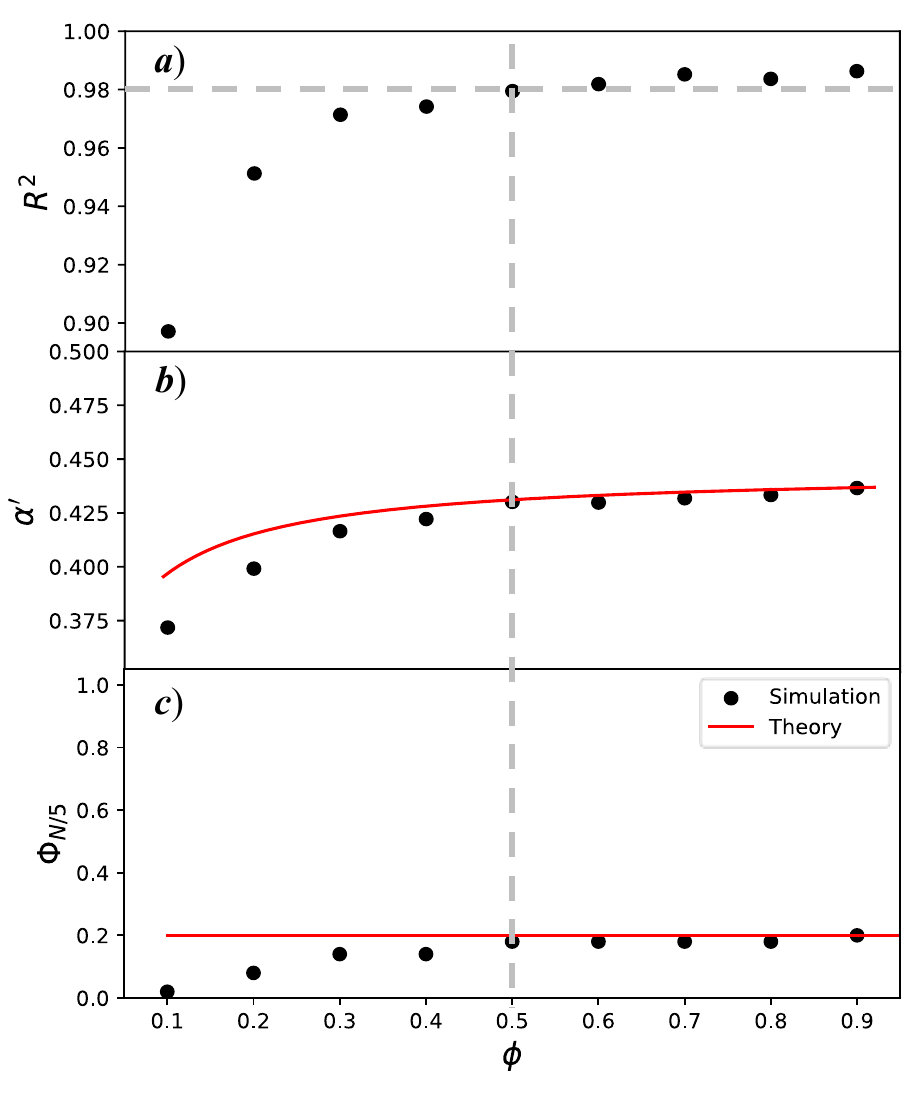}
    \caption{\textbf{The critical $R^*$ is sufficient to evaluate the effectiveness of the approximation.}
    (a) The coefficient of determination, $R^2$, increases with $\phi$. For $\phi>\phi^*$, the theoretical solutions of (b) $\alpha^{\prime}$ and (c) fixation probability coincide with the simulation results, which validates Ansatz~\ref{Ansatz1}. The simulation data points are calculated by averaging over $50$ independent runs. For every realization, the population starts with $20$ individuals with opinion $A$, and all the simulation data points are calculated by averaging over $50$ independent runs. Here $N=100$, $L=400$, $k_d=0.6$ and $k_s=0.3$. The two dashed lines represent $R^*$ and $\phi^*$ respectively, where $R^*=0.98$ and $\phi^*=0.5$.}
    \label{fig: rap}
\end{figure}

    We find that $R^2$ increases with $\phi$, the proof of which is provided in Appendix \ref{Appendix: R} (see Fig.~\ref{fig: rap}$a$ for simulation). Given the breaking edge probability and the network configuration, there exists a $\phi^*$ corresponding to $R^*$. If $\phi>\phi^*$ ($R^2>R^*$), then $v=w$ holds. Bringing the equation $v=w$ into the nonlinear system Eqs. (\ref{eq:4}-\ref{eq:6}), another one-dimensional manifold emerges given by (See Appendix \ref{Appendix: general_manifold})
    \begin{numcases}{}
        u = 1-\alpha^\prime(1-w^2),\label{eq:13}\\
        v = w,\label{eq:14}
    \end{numcases}
    where $\alpha^\prime=\frac{1}{4}-\frac{1}{4\bar{k}}-\frac{(1-\phi)(k_d+k_s)}{8\phi\bar{k}}+\frac{\sqrt{\Delta_0}}{8\phi\bar{k}}$ and $\Delta_0=[2\phi(3\bar{k}+1)+(1-\phi)(k_d+k_s)]^2-16\phi\bar{k}[2\phi(\bar{k}+1)+(1-\phi)k_d]$.

    Eqs. (\ref{eq:13}-\ref{eq:14}) are the approximated stable manifold for $k_d\neq k_s$.
    Note-worthily, the emergent manifold bears the same form as that of the benchmark case Eqs. (\ref{eq:7}-\ref{eq:8}) with a modified $\alpha\prime$. If $k_d=k_s=k_0$, then the one-dimensional manifold Eqs. (\ref{eq:13}-\ref{eq:14}) degenerates to that of the benchmark case.


    When $\phi>\phi^*$, the theoretical solutions Eqs. (\ref{eq:13}-\ref{eq:14}) of the manifold are confirmed by simulation in Fig.~\ref{fig: rap}$b$, which validates Ansatz~\ref{Ansatz1}. The introduction of $R^2$ is thus an effective way to measure the validity of the approximated manifold.     
    When $\phi<\phi^*$, the difference between simulation and theoretical results is non-negligible. It implies that the accuracy of the theoretical solution is closely related to $R^2$. In other words, Ansatz~\ref{Ansatz1} is no longer valid for $R^2<R^*$. Therefore, a new ansatz needs to be proposed to overcome the inaccuracy of Ansatz~\ref{Ansatz1}.

    \begin{ansatz}
    $v(w;\phi)$ satisfies the convex combination of two extremes with respect to $\phi$, namely
    \begin{align}
        v(w;\phi)=(1-\phi)\cdot v(w;0^+)+\phi \cdot v(w;1^-),\label{eq:anz}
    \end{align}\label{Ansatz2}
    where $v(w;\phi_0)$ denotes the relationship between $v$ and $w$ when $\phi=\phi_0$.
    \end{ansatz}

    If $\phi\to 1^-$, only Process $2$ happens and our model becomes the voter model on static networks. At this point, our model is independent of $k_d$ and $k_s$. The manifold (\ref{eq:7}, \ref{eq:8}) emerges and $v=w$ holds. Thus, 
    \begin{align}
        v(w;1^-)=w.\label{eq:v1}
    \end{align}

    If $\phi\to 0^+$, only Process $1$ happens and the proportions of opinions remain unchanged. At this point, the proportions of various edges have a unique stationary distribution \cite{wu2010evolution} $(\frac{[AA]}{L},\frac{[AB]}{L},\frac{[BB]}{L})=\mathcal{N}([A])(\frac{[A]^2}{k_s},\frac{2[A][B]}{k_d},\frac{[B]^2}{k_s})$, where $\mathcal{N}([A])=(\frac{[A]^2}{k_s}+\frac{2[A][B]}{k_d}+\frac{[B]^2}{k_s})^{-1}$ is the normalization factor. The relationship between $v$ and $w$ is given by 
    \begin{align}
        v(w;0^+)&=\frac{[AA]}{L}-\frac{[BB]}{L}\nonumber\\
        &=\frac{w}{ew^2+(1-e)},\label{eq:v2}
    \end{align}
    where $e=(1-{k_s}/{k_d})/2$.

    We make a perturbation to the two probabilities of breaking edges of the benchmark case. Specifically, the two probabilities of breaking edges are given by $k_d=k_0$ and $k_s=k_0-\theta\delta$,    
    where $\delta\to 0^+$ and it represents a slight perturbation. And $\theta$ indicates the magnitude of the perturbation. If $\theta>0$, i.e., $k_d>k_s$, then the population is of in-group bias. If $\theta<0$, i.e., $k_d<k_s$, then the population is of out-group bias.

    Bringing Eqs. (\ref{eq:v1}-\ref{eq:v2}) into Eq.~(\ref{eq:anz}), we obtain the modified ansatz,
    \begin{align}
        v(w;\phi)
        = \! w\!+\!(1\!-\!\phi)\frac{\theta}{2k_0}w(1\!-\!w)(1\!+\!w)\delta\!+\!o(\delta).\label{eq:anz2}
    \end{align}

    We take Ansatz~\ref{Ansatz2} (i.e., Eq.~\eqref{eq:anz2}) as the stable manifold for $k_d\neq k_s$. 
    As $\phi$ increases, the absolute value of the second term of Eq.~\eqref{eq:anz2} decreases, which indirectly explains $R^2$ increases with $\phi$ (The proof is provided in Appendix \ref{Appendix: R}). 

\subsubsection{Fixation probability}

The stable manifold also facilitates us to capture the fate of the system. If $R^2>R^*$, the manifold is represented well by Eqs. (\ref{eq:13}-\ref{eq:14}) and the motion on the manifold is also a diffusion process with neutral drift, thus the fixation probability is $\Phi_j=j/N$.

$R^2>R^*$ means the gap between the real manifold and the approximated manifold $v=w$ is negligible. It also means $\phi$ needs to be sufficiently large. As shown in Fig.~\ref{fig: rap}$c$, $\phi^*$ is around $0.5$ and $\Phi_j=j/N$ holds well when $\phi>\phi^*$.
The result shows that the fixation probability $\Phi_j=j/N$ of the benchmark case (i.e., $k_d=k_s$) is robust to the general case (i.e., $k_d\neq k_s$) if $\phi>\phi^*$. The manifold of the benchmark case is thus robust to capture the fate of the system (i.e., the fixation probability).

\begin{figure}[hbtp]
    \centering
    \includegraphics[width=\columnwidth]{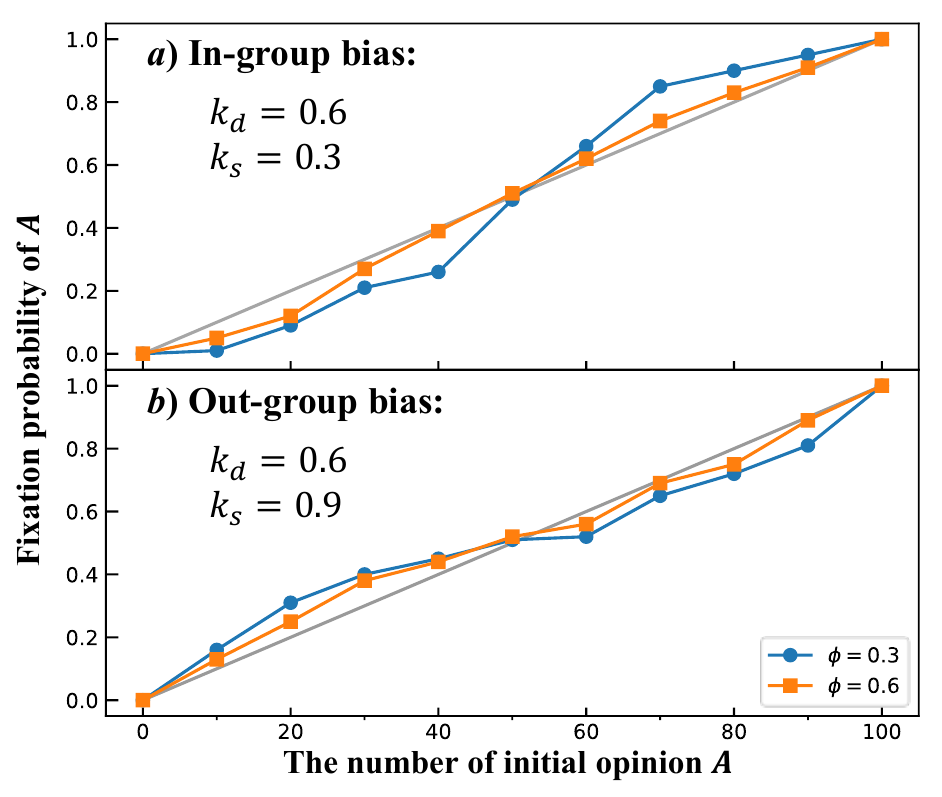}
    \caption{\textbf{The fixation probability for different social biases can be interpreted from an evolutionary game perspective.}
    (a) In-group bias and (b) out-group bias. The circle dots represent the simulation results for $\phi=0.3$ while the square dots represent those of $\phi=0.6$. Compared with the benchmark case, for in-group bias, it's harder for the minority opinion to take over the population while it's easier for the majority opinion. 
    In a population with in-group bias, conformity inhibits the domination of the majority opinion but promotes the invasion of the minority opinion. The results agree with Eq.~(\ref{eq:fb}), which validates Ansatz~\ref{Ansatz2}. 
    Besides, the curves are similar to Fig.~1 in Ref.~\onlinecite{traulsen2006stochastic}, which implies that from a fixation probability perspective, the equivalence between opinion dynamics and the evolutionary game, which Ref.~\onlinecite{wu2020CCC} presents, is somehow robust for arbitrary $\phi$.}
    \label{fig: FP}
\end{figure}

If $R^2<R^*$, the motion on the one-dimensional manifold is no longer neutral drift because $v=w$ doesn't hold. The transition probabilities become

\begin{numcases}{}
    P_j^+=\phi\cdot\frac{1-w}{2}\cdot\frac{1-u}{1-v(w;\phi)}\bigg|_{w=\frac{2j}{N}-1},\\
    P_j^-=\phi\cdot\frac{1+w}{2}\cdot\frac{1-u}{1+v(w;\phi)}\bigg|_{w=\frac{2j}{N}-1}.
\end{numcases}
where $v(w;\phi)$ is given by Ansatz~\ref{Ansatz2} (Eq.~\eqref{eq:anz2}).

    Based on the one-dimensional Markov chain, the fixation probability $\Phi_j$ can be given by (See Appendix \ref{Appendix: perturbation_fp}),
    \begin{align}
        \!\frac{j}{N}\!+\!\frac{(1\!-\!\phi)j}{6N^2k_0}\theta\underbrace{\left[(j\!-\!1)(3N\!-\!2j\!-\!2)\!-\!(N\!-\!1)(N\!-\!2)\right]}_\Gamma\delta\!+\!o(\delta).\label{eq:fb}
    \end{align}

    The first term of Eq.~(\ref{eq:fb}) is the fixation probability of the benchmark case.
    The sign of the second term of Eq.~\ref{eq:fb}) is determined by $\theta$ and $\Gamma$. Herein, $\theta$ is determined by social bias and $\Gamma$ is determined by $j$. If $j<N/2$, then $\Gamma<0$. If $N/2<j\le N$, then $\Gamma>0$.

    For $\theta>0$, if $j<N/2$ the second term of Eq.~\eqref{eq:fb} is negative while if $j>N/2$ it is positive.
    It implies that for in-group bias it is harder for only one individual with opinion $A$ to take over the entire population than that of the benchmark case. However, once the proportion of opinion $A$ is more than one-half, it becomes easier to take over the entire population than that of the benchmark case. It is intuitive. A population with in-group bias doesn't tolerate a new and innovative opinion. Therefore, it is more challenging for a new opinion to take over the entire population with in-group bias than that of the benchmark case.

    The results can also be interpreted from the perspective of evolutionary game theory. In fact, the rewiring probability, i.e., $(1-\phi)$, is a rescaling of the selection intensity in evolutionary games. The equivalence between opinion dynamics and game theory was first proposed in Ref.~\onlinecite{wu2020CCC} but it is only applied to $\phi\to 0^+$ therein.  
    As shown in Fig.~\ref{fig: FP}, for any $\phi\in(0,1)$, the fixation probability for in-group bias coincides with that of the Moran process, whose payoff matrix is a coordination game \cite{traulsen2006stochastic}.
    It implies that the equivalence is robust for $0<\phi<1$.
    Meanwhile, the opposite is true for out-group bias ($k_d<k_s$). 

    For $\theta>0$, if $j<N/2$ the fixation probability $\Phi_j$ increases with $\phi$ whereas if $j>N/2$ the fixation probability $\Phi_j$ decreases with $\phi$.
    It means that for in-group bias if the individuals are more likely to adjust their opinions (i.e., $\phi$ gets larger), it's easier for the minority opinion to dominate the entire population, but it's harder for the majority opinion to dominate it. It is counterintuitive. The parameter $\phi$ denotes the occurrence probability of opinion propagation. In other words, it is an indicator of the degree of conformity in a population, because the logic behind adjusting opinions (i.e., Process $2$) is based on conformity. When $\phi$ gets larger, individuals are less confident of their own opinions and are more likely to adopt the popular opinion. Therefore, conformity favors the majority opinion intuitively.
    Besides, it is also intuitive that under in-group bias, the majority opinion takes advantage too because individuals in such a population voluntarily stay away from those with strange and peculiar opinions, i.e., the minority opinions.
    However, as conformity enhances (i.e., $\phi$ increases), it becomes more difficult for the majority opinion to take over the entire population with in-group bias. It contradicts both the conformity and in-group bias.
    For out-group bias, the opposite is true. If the nodes are more likely to adjust their opinions, it's harder for the minority opinion to dominate the entire population, but it's easier for the majority opinion to dominate it.

If the nodes are more likely to adjust their social relationships (i.e., as $\phi$ decreases), the difference in the fixation probabilities between the general case and the benchmark case will be more pronounced. In other words, the robustness of the probability of the benchmark case increases with $\phi$. It explains that $\Phi_j=j/N$ holds well when $\phi>\phi^*$ from another perspective.

\subsubsection{Fixation time}

    \begin{figure}[htbp]
        \centering
        \includegraphics[width=\columnwidth]{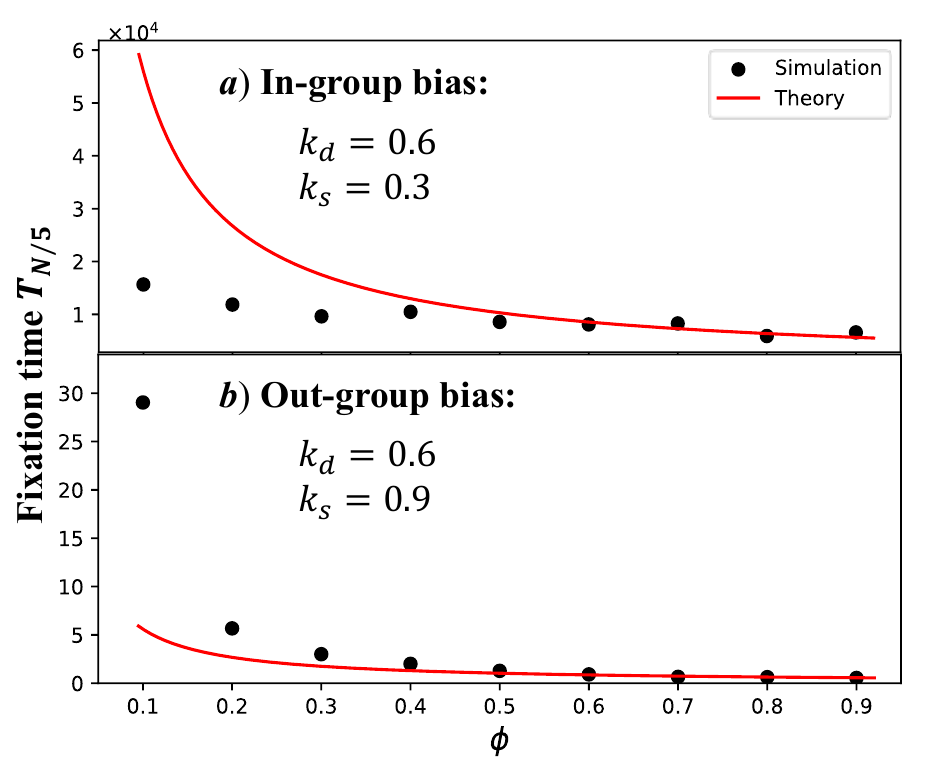}
        \caption{\textbf{It takes a short time for in-group bias to reach consensus whereas it takes a long time for out-group bias.}
        If $\phi>\phi^*$, the theoretical solution Eq.~(\ref{eq:B19}) of fixation time coincides with the simulation results well. Otherwise, the simulation results are lower for in-group bias while the simulation results are higher for out-group bias. It implies that from a fixation time perspective, the equivalence between opinion dynamics and the evolutionary game is somehow robust for arbitrary $\phi$.}
        \label{fig: ft}
    \end{figure}

    The fixation time of the benchmark case is also robust to that of the general case. In contrast with the benchmark case, it is only valid for $R^2>R^*$ (i.e., $\phi \ge \phi^*$) as shown in Fig.~\ref{fig: ft}.

    As shown in Fig.~\ref{fig: ft}, compared with the theoretical values, the simulation results are lower for in-group bias and higher for out-group bias if $\phi<\phi^*$. We find that it can be interpreted from an evolutionary game perspective. Ref.~\onlinecite{antal2006fixation} and Ref.~\onlinecite{wu2020CCC} together demonstrate that if $\phi \to 0^+$, compared with the benchmark case, it takes a longer time for in-group bias to reach fixation (consensus) while it takes a shorter time for out-group bias. Because if $\phi\to 0^+$, the in-group (out-group) bias corresponds to the coordination (co-existence) game \cite{wu2020CCC} and the fixation time of the coordination (co-existence) game is smaller (larger) than that under neutral selection \cite{antal2006fixation,altrock2009fixation}. The results validate that the correspondence between opinion dynamics and the evolutionary game is not only valid if $\phi\to 0^+$, but also somehow robust for arbitrary $\phi$.

\section{Discussion and Conclusion}

We have modeled the social bias without introducing highly cognitive ability of individuals.
On one hand, our model is able to capture social bias by heterogeneous breaking probabilities across social ties. 
If the $AB$ edges are more likely to break off, then the population is of in-group bias. Otherwise, the population is of out-group bias.
While taking into social bias account, we have not ignored that any social connection is not perfectly strong in reality. All social relationships in our model are likely to break off, which is realistic.
On the other hand, individuals in our model rewire to a random individual as their new neighbor when updating their social relationships.
This implies that individuals aren't necessarily to know the opinions of others before rewiring.
Individuals in previous works typically either rewire to those with the same opinion or to those with different opinions \cite{holme2006nonequilibrium,kimura2008coevolutionary,simplex2019}.
Although the rewire-to-same differs from rewire-to-different, the implicit assumptions in both models are the same: all the individuals know all the opinions of all the rest in the population.
This is unrealistic because individuals' cognitive abilities are overestimated.
Individuals in our models thus have limited cognitive ability.
In addition, the rewire-to-same may lead to fragmentation \cite{holme2006nonequilibrium,kimura2008coevolutionary}, which is unrealistic in the information and interconnected age.

Opinion dynamics on dynamical networks are typically based on simulations or numerical solutions \cite{holme2006nonequilibrium,simplex2019,kozma2008consensus,zimmermann2004coevolution,oestereich2023optimal,durrett2012graph,du2023coevolutionary,fu2008coevolutionary}. Existing theoretical analysis are not large in number compared with the simulation results. Theoretical approaches to tackle the coevolutionary dynamics often assume that social interactions evolve much faster than opinions, i.e., sufficiently large rewiring probability \cite{wu2010evolution,wu2011evolutionary,wu2016evolving,wu2019evolution,wu2020CCC,wei2019vaccination,shan2022social,wang2023opinions,liu2023emergence,baumann2020modeling,AT2006activelink,du2022evolutionary,du2023coevolutionary,du2024asymmetric}. In this way, the network structure has converged to a steady state whenever the opinion is updated.
We extrapolate from this assumption and explicitly address how opinions reach consensus for arbitrary rewiring probability. 
When the probabilities of breaking edges are equal, i.e., the benchmark case, we use the pair-approximation and the mean-field equation to capture the system.
And we find a central manifold and prove the stability of the central manifold.
It turns out to be the most likely trajectory.
For the manifold, we give a theoretical solution, instead of fitting parabolas\cite{simplex2019} or numerical solutions of the approximate master equations\cite{durrett2012graph,marceau2010adaptive}.
When the probabilities of breaking edges are not equal, we introduce for the first time the coefficient of determination, $R^2$.
We find that it provides a theoretical basis for the approximated stable manifold.
Therefore, we find the most likely trajectory of the coevolutionary dynamics, which was not achievable by previous works.
In addition, we provide a modified ansatz to approximate the most likely trajectory for general cases.

We investigate both the fixation probability and the fixation time with the aid of the most likely trajectory. 
We find that for the in-group bias, it's harder for the minority opinion to invade than that of the benchmark case, but it's easier for the majority opinion to dominate the population than that of the benchmark case. In addition, we find that it takes a shorter time to reach the consensus for in-group bias but a longer time for out-group bias.
Furthermore, we are able to give a game-theoretical explanation on all these findings:
Ref.~\onlinecite{wu2020CCC} presents that if the network evolves much faster than opinions updating (i.e., sufficiently large rewiring probability), the evolution of the opinion ratios in the populations with in-group bias (out-group bias) can be equivalent to the Moran process with a payoff matrix of the coordination (co-existence) game. 
Here we find that the equivalence between opinion dynamics and evolutionary game theory is not only valid for sufficiently large rewiring probability, but also somewhat robust for arbitrary rewiring probability. 
The above conclusions about both the fixation probability \cite{traulsen2006stochastic} and the fixation time \cite{antal2006fixation,altrock2009fixation} agree with the equivalence.

Our work investigates a binary, one-dimensional opinion model and verifies the theoretical results in terms of both fixation probability and fixation time. The theoretical method in the paper can be promising for high-dimensional opinion models \cite{fu2018codiffusion}. 
In addition, the echo-chamber has become the focus of current research topics \cite{fu2018opinion,wang2020public,liu2023emergence,baumann2020modeling,flamino2023political,cinelli2021echo,steiglechner2023social}.
The fixation time measures the time of co-existence of opinions, which also indirectly measures the echo-chamber effect \cite{vvv2019consensus,borges2024social}. It can be promising in the future. 

To sum up, with the aid of a novel approximation method for any opinion updating frequency,
we find that 
conformity inhibits the majority opinion from taking over the in-group biased population on dynamical networks.

\begin{acknowledgments}
We are grateful for the discussions with Yakun Wang. This work is supported by National Natural Science Foundation of China (NSFC) under Grant 61751301.
\end{acknowledgments}

\section*{Author DECLARATIONS}
\subsection*{Conflict of Interest}
The authors have no conflicts to disclose.
\subsection*{Author Contributions}

\section*{Data Availability Statement}


The data that supports the findings of this study are openly
available in GitHub at https://github.com/wxl-001/adaptive.

\appendix

\section{Benchmark case: $k_d=k_s=k_0$\label{Appendix: Benchmark case}}

In this section, let us consider a benchmark case, where $k_d=k_s=k_0$.

\subsection{The stable manifold\label{Appendix:manifold}}

Applying $k_d=k_s=k_0$ to Eqs(\ref{eq:4}-\ref{eq:6}), we obtain
\begin{align}
    \frac{du}{dt}=&\phi\frac{2}{N\bar{k}}\cdot\frac{1-u}{1-v^2}\left[1-vw+\bar{k}(1-2u+v^2)\right]\nonumber\\&+\frac{1-\phi}{N\bar{k}}\cdot k_0(1-2u+vw),\label{eq:B1}\\
    \frac{dv}{dt}=&\frac{v-w}{N\bar{k}}\left[2\phi\cdot\frac{1-u}{1-v^2}-(1-\phi)k_0\right],\label{eq:B2}\\
    \frac{dw}{dt}=&\phi\frac{2}{N}\cdot\frac{1-u}{1-v^2}\cdot(v-w).\label{eq:B3}
\end{align}

The Eqs(\ref{eq:B1}-\ref{eq:B3}) have a set of fixed points on
    \begin{numcases}{}
        u = 1-\alpha(1-w^2),\label{eq:B4}\\
        v = w,\label{eq:B5}
    \end{numcases}

\noindent where $\alpha=\frac{\phi(\bar{k}\!-\!1)\!-\!(1\!-\!\phi)k_0+\sqrt{k_0^2(1\!-\!\phi)^2\!+\!2(1\!+\!\bar{k})k_0\phi(1\!-\!\phi)\!+\!\phi^2(\bar{k}\!-\!1)^2}}{4\phi\bar{k}}$. Surprisingly, any fixed point can be represented by a variable $w$. The process of solving the manifold (\ref{eq:B4},\ref{eq:B5}) is provided in Appendix \ref{Appendix: general_manifold}.

Next, we prove that the set of fixed points is a one-dimensional stable manifold on the three-dimensional space.

Take any one point on the set of fixed points. Denote it as $p_1$, where $p_1=[u_1,v_1,w_1]^T=[1-\alpha(1-{w_1}^2),w_1,w_1]^T$. Linearizing the system at point $p_1$, we obtain 
\begin{equation}\label{eq:B6}
\frac{d[u,v,w]^T}{dt}\Bigg|_{p_1}=
\left[
\begin{array}{ccc}
    d & aw_1 & bw_1\\
    0 & -b & b\\
    0 & c & -c
\end{array}
\right]\cdot
\left[
\begin{array}{c}
     u \\
     v\\
     w
\end{array}
\right].
\end{equation}

We have used the notations
\begin{numcases}{}
    a=\frac{2\phi\alpha(1+4\bar{k}\alpha)+k_0(1-\phi)}{N\bar{k}},\nonumber\\
    b=\frac{k_0(1-\phi)-2\phi\alpha}{N\bar{k}},\nonumber\\
    c=\frac{2\phi\alpha}{N},\nonumber\\
    d=-\frac{2\sqrt{k_0^2(1-\phi)^2+2(1+\bar{k})k_0\phi(1-\phi)+\phi^2(\bar{k}-1)^2}}{N\bar{k}}\nonumber,
\end{numcases}
 where notations $a$, $b$ and $c$ are all positive and notation $d$ is negative when $\bar{k}\ge 2$. It must be a disconnected network for $\bar{k}<2$, so we do not consider this case.

 The eigenvalues at the set of fixed points (\ref{eq:B4}-\ref{eq:B5}) are
\begin{align}
    \lambda=
    \left\{
    \begin{array}{l}
         0 \ , \\
         d \ ,\\
         -b-c \ .
    \end{array}\label{eq:B11}
    \right.
\end{align}

Since only one of the eigenvalues(\ref{eq:B11}) is 0 and the rest are negative, thus the set of fixed points(\ref{eq:B4}-\ref{eq:B5}) is a one-dimensional stable manifold based on the center manifold theorem\cite{strogatz2018nonlinear,Khalil2008NonlinearST}.

\subsection{Fixation Probability and Fixation Time\label{Appendix: fixation}}

    From Appendix \ref{Appendix:manifold} we know that the three-dimensional system converges quickly to a one-dimensional slow manifold. We can represent the motion on the slow manifold by a one-dimensional Markov chain with $[A]$ (the Markov chain is well defined in the main text).

    The state of the Markov chain is influenced only by Process $2$ and only transfers to adjacent states in one step. $P^{+}_j$ ($P^{-}_j$) denotes the probability that the Markov chain increases (decreases) by one from state $j$,
    \begin{align}
        P^+_j&=\phi\cdot\frac{[B]}{N}\cdot R(A|B)\nonumber\\
            &=\phi\cdot\frac{[B]}{N}\cdot\frac{[AB]}{[AB]+2[BB]}\nonumber\\
            &=\phi\cdot\frac{1-w}{2}\cdot\frac{1-u}{1-v}\bigg|_{w=\frac{2j}{N}-1}\nonumber\\
            &\xlongequal[v=w]{u=1-\alpha(1-w^2)} \frac{\alpha\phi}{2}(1-w^2)\bigg|_{w=\frac{2j}{N}-1}\nonumber\\
            &= \frac{\alpha\phi}{2}[1-(\frac{2j}{N}-1)^2].\label{eq:B12}
    \end{align}

    Similarly, we obtain $P^-_j=\frac{\alpha\phi}{2}[1-(\frac{2j}{N}-1)^2]$. Since $P^+_j=P^-_j$, the motion on the one-dimensional Markov chain is the diffusion process of neutral drift. Fixation probability, $\Phi_j$, with initial state $j$ is given by $j/N$\cite{traulsen2009stochastic,Gardiner1986HandbookOS}.

    Ref.~\onlinecite{traulsen2009stochastic} presented the formula of fixation time and invasion time,
    \begin{numcases}{}
     T_j=-T_1\sum_{k=j}^{N-1}\prod_{m=1}^k\frac{P^-_j}{P^+_j}+\sum_{k=j}^{N-1}\sum_{l=1}^k\frac1{P_l^+}\prod_{m=l+1}^k\frac{P^-_j}{P^+_j},\label{eq:B13}\\
     T_1=\frac1{1+\sum_{k=1}^{N-1}\prod_{j=1}^k\frac{P^-_j}{P^+_j}}\sum_{k=1}^{N-1}\sum_{l=1}^k\frac1{P_l^+}\prod_{j=l+1}^k\frac{P^-_j}{P^+_j}.\label{eq:B14}
    \end{numcases}

    Based on the formula of invasion time (\ref{eq:B14}), we obtain 
    \begin{align}
        T_1=&\frac{1}{N}\sum_{k=1}^{N-1}\sum_{l=1}^k\frac{1}{P^+_l}\nonumber\\
        =&\frac{1}{N}\sum_{k=1}^{N-1}\sum_{l=1}^k\frac{1}{\alpha\phi}\cdot\frac{2}{1-w^2}\bigg|_{w=\frac{2j}{N}-1}\nonumber\\
        =&\frac{1}{N}\sum_{k=1}^{N-1}\sum_{l=1}^k\frac{1}{\alpha\phi}\left(\frac{1}{1-w}+\frac{1}{1+w}\right)\bigg|_{w=\frac{2j}{N}-1}\nonumber\\
        =&\frac{1}{N}\sum_{k=1}^{N-1}\sum_{l=1}^k\frac{N}{2\alpha\phi}\left(\frac{1}{N-l}+\frac{1}{l}\right)\nonumber\\
        =&\frac{1}{2\alpha\phi}\sum_{k=1}^{N-1}\sum_{l=1}^k\left(\frac{1}{N-l}+\frac{1}{l}\right).
    \end{align}

   The sums in the previous equation can be interpreted as numerical approximations to the integrals (i.e., $\sum_{k=1}^i\ldots\approx\int_1^i\ldots dk$)\cite{traulsen2006stochastic,traulsen2009stochastic}. Replacing the sums with the integrals, we obtain 
   \begin{align}
       T_1=&\frac{1}{2\alpha\phi}\int_1^{N-1}\left[\int_1^k \left(\frac{1}{N-l}+\frac{1}{l}\right)dl\right]dk\nonumber\\
       =&\frac{1}{2\alpha\phi}(N-1)\ln(N-1).\label{eq:B16}
   \end{align}

   The result is the invasion time, i.e., the fixation time starts from only an opinion $A$. Similarly, we also obtain the formula of $T_j$,

    \begin{align}
       T_j=&\frac{N}{2\alpha\phi}[(2N-j-1)\ln (N-1)-j\ln j\nonumber\\
       &-(N-j)\ln (N-j)]-(N-j)T_1\label{eq:B17}\\
       =& \frac{1}{2\alpha\phi}\{(N^2-j)\ln(N-1)\nonumber\\
       &-N[(N-j)\ln(N-j)+j\ln j]\}.\label{eq:B18}
   \end{align}

   Eq.~(\ref{eq:B18}) is obtained by bringing Eq.~(\ref{eq:B16}) into Eq.~\ref{eq:B17}) but the symmetry (i.e., $T_j=T_{N-j}$) cannot be satisfied. For this reason, we make a small change to Eq.~(\ref{eq:B16}), $T_1=N\ln (N-1)/(2\alpha\phi)$, and bring it into Eq.~(\ref{eq:B17}). $T_j$ is approximated by
   \begin{align}
       \frac{N}{2\alpha\phi}\left\{(N-1)\ln(N-1)-[(N-j)\ln(N-j)+j\ln j]\right\}.\label{eq:B19}
   \end{align}

    Eq.~(\ref{eq:B19}) ensures the symmetry of fixation time. Based on Eq.~(\ref{eq:B19}), we find that if $j=N/2$ (assuming $N$ is an even number), the fixation time $T_j$ reaches its maximum and it is of $O(N^2)$.
    \begin{align}
        T_{\frac{N}{2}}&=\frac{1}{2\alpha\phi}\left[(N^2-N)\ln(N-1)-N^2\ln\frac{N}{2}\right]\label{eq:B20}\\
        &=\frac{1}{2\alpha\phi}\left[N^2\ln\frac{2N-2}{N}-N\ln (N-1)\right]\nonumber\\
        &=\frac{N^2}{2\alpha\phi}\left[\ln(2-\frac{2}{N})-\frac{\ln(N-1)}{N}\right]\nonumber\\
        &\xrightarrow{N\to \infty}\frac{\ln2}{2\alpha\phi}N^2.\label{eq:B21}
    \end{align}

\section{The manifold of the general case\label{Appendix: general_manifold}}
If $R^2>R^*$, then we assume that $v=w$ holds well. Bringing it into the three-dimensional nonlinear system Eqs. (\ref{eq:4}-\ref{eq:6}), we obtain
\begin{align}
    \frac{du}{dt}=&\frac{2\phi}{N\bar{k}}\frac{1-u}{1-w^2}[1-w^2+\bar{k}(1-2u+w^2)]\nonumber\\&+\frac{1-\phi}{N\bar{k}}[k_d(1-u)-k_s(u-w^2)],\label{eq:C1}\\
    \frac{dw}{dt}=&\frac{dv}{dt}=0.\label{eq:C2}
\end{align}

 The right hand of Eq.~(\ref{eq:C1}) equivalent to $0$, namely Eq.~\ref{eq:C3}), is an implicit function expression for the manifold. 
\begin{widetext}
    \begin{align}
    &G(u;\phi,k_d,k_s,\bar{k},w)\nonumber\\
    =&2\phi(1-u)\left[1-w^2+\bar{k}(1-2u+w^2)\right]+(1-\phi)(1-w^2)\left[k_d(1-u)-k_s(u-w^2)\right]\label{eq:C3}\\
    =&4\bar{k}\phi u^2-[2\phi(3\bar{k}+\bar{k}w^2-w^2+1)+(1-\phi)(k_d+k_s)(1-w^2)]u+2\phi[1-w^2+\bar{k}(1+w^2)]+(1-\phi)(1-w^2)(k_d+k_sw^2)\nonumber\\
    =&0.\nonumber
\end{align}
\end{widetext}

$G(u;\phi,k_d,k_s,\bar{k},w)$ is a quadratic function about $u$, where $G(0)>0$ and $G(1)=-k_s(1-\phi)(1-w^2)^2<0$. Thus, for any $w$, a value of $u$ between $0$ and $1$ must exist, which is the smaller of the two roots of the quadratic equation (\ref{eq:C3}). It is denoted by $u^*$ and is determined by $\phi$, $\bar{k}$, $k_d$, $k_s$ and $w$. The discriminant of the quadratic equation (\ref{eq:C3}) is denoted by $\Delta$, where $\Delta=\Delta_0(1-w^2)^2$ and $\Delta_0=[2\phi(\bar{k}-1)-k_d(1-\phi)]^2+[k_s(1-\phi)]^2+4\phi(1-\phi)(3\bar{k}+1)k_s+2k_dk_s(1-\phi)^2$. Based on the quadratic formula, $u^*$ is given by 
\begin{align}
    u^*&\!=\!\frac{2\phi(3\bar{k}\!+\!\bar{k}w^2\!-\!w^2\!+\! 1)\!+\!(1\!-\!\phi)(k_d\!+\!k_s)(1\!-\!w^2)\!-\!\Delta}{8\phi\bar{k}}\nonumber\\
    &\!=\! 1\!-\!\underbrace{\left[\frac{1}{4}\!-\!\frac{1}{4\bar{k}}\!-\!\frac{(1\!-\!\phi)(k_d\!+\!k_s)}{8\phi\bar{k}}\!+\!\frac{\sqrt{\Delta_0}}{8\phi\bar{k}}\right]}_{\alpha^\prime}(1\!-\!w^2).\label{eq:C4}
\end{align}

Eqs. (\ref{eq:13}-\ref{eq:14}) in the main text are well proven by Eq.~\ref{eq:C4}). Besides, the manifold (\ref{eq:B4}, \ref{eq:B5}) of the benchmark case is obtained in the same way. If $k_d=k_s=k_0$, the manifold of the general case degenerates to that of the benchmark case.

Eq.~(\ref{eq:C2}) implies that the motion on the manifold is the one-dimensional diffusion process of neutral drift, i.e., random walk on the manifold. The transition probabilities on the one-dimensional manifold are still represented by Eq.~(\ref{eq:B12}). Thus, if $R^2>R^*$, the fixation probability $\Phi_j=j/N$ and the fixation time Eqs. (\ref{eq:B16}-\ref{eq:B21}) of the benchmark case are robust to the general case. Herein, $\alpha$ is replaced by $\alpha^\prime$.

\section{$R^2$ increases with $\phi$\label{Appendix: R}}
We consider the perturbed $v$ (Eq.~\eqref{eq:anz2}) as the original dataset. We consider $v=w$ as the fitting function and introduce the coefficient of determination, $R^2=1-\sum(y_i-\hat{y})^2/\sum(y_i-\bar{y})^2$ to measure the goodness of fit of the linear regression model. \cite{dunn2018generalized,mccullagh1989generalized}. At this point, $R^2$ is given by
\begin{align}
    R^2=&1-\frac{\sum_{w\in (-1,1)}(v(w;\phi)-w)^2}{\sum_{w\in (-1,1)}(v(w;\phi)-0)^2}\nonumber\\
    =&1-\frac{\int_0^1\left[(1-\phi)\cdot\frac{\theta}{2k_0}w(1-w)(1+w)\cdot \delta\right]^2dw}{\int_0^1\left[w+(1-\phi)\cdot\frac{\theta}{2k_0}w(1-w)(1+w)\cdot \delta\right]^2dw}\nonumber\\
    =&1-\frac{\frac{2}{105}\left[\frac{(1-\phi)\theta}{k_0}\right]^2\cdot\delta^2}{\frac{1}{3}+\frac{2(1-\phi)\theta}{15k_0}\delta+\frac{2}{105}\left[\frac{(1-\phi)\theta}{k_0}\right]^2\delta^2}\nonumber\\
    =&1-\frac{4}{35}\left[\frac{(1-\phi)\theta}{k_0}\right]^2\delta^2+o(\delta^2).\label{eq:D6}
\end{align}
Eq.~(\ref{eq:D6}) shows that $R^2$ increases with $\phi$.

\section{Perturbation on the fixation probability\label{Appendix: perturbation_fp}}
Without taking into account the assumption that $v=w$, the motion on the one-dimensional manifold is no longer neutral drift. The transition probabilities become 
\begin{numcases}{}
    P_j^+=\phi\cdot\frac{1-w}{2}\cdot\frac{1-u}{1-v}\bigg|_{w=\frac{2j}{N}-1},\\
    P_j^-=\phi\cdot\frac{1+w}{2}\cdot\frac{1-u}{1+v}\bigg|_{w=\frac{2j}{N}-1}.
\end{numcases}

The fixation probability $\Phi_j$ with initial state $j$ satisfies the following equation \cite{traulsen2009stochastic,Gardiner1986HandbookOS}:
\begin{align}
    \Phi_j=\frac{1+\sum_{k=1}^{j-1}\Pi_{j=1}^k\gamma_j}{1+\sum_{k=1}^{N-1}\Pi_{j=1}^k\gamma_j},\label{eq:D9}
\end{align}
where 
\begin{align}
    \gamma_j=\frac{P_j^-}{P_j^+}=\frac{1+w}{1+v}\cdot\frac{1-v}{1-w}\bigg|_{w=\frac{2j}{N}-1}.\label{eq:D10}
\end{align}

Bringing the perturbed $v$ into Eq.~\ref{eq:D10}),
\begin{align}
    \gamma_j=&\frac{1-w^2-(1-\phi)\frac{\theta}{2k_0}w(1-w)(1+w)^2\delta}{1-w^2+(1-\phi)\frac{\theta}{2k_0}w(1-w)^2(1+w)\delta}\bigg|_{w=\frac{2j}{N}-1}\nonumber\\
    \approx&1-(1-\phi)\frac{\theta w}{k_0}\cdot \delta\bigg|_{w=\frac{2j}{N}-1}.
\end{align}

The product in Eq.~(\ref{eq:D9}) can then be simplified to
\begin{align}
    \prod_{j=1}^k\gamma_j=&\prod_{j=1}^k\left[1-(1-\phi)\frac{\theta w}{k_0}\cdot\delta\right]\bigg|_{\frac{2j}{N}-1}\nonumber\\
    \approx&1-\delta\sum\limits_{j=1}^k\left[(1-\phi)\frac{\theta}{k_0}(\frac{2j}{N}-1)\right]\nonumber\\
    =&1-\delta[\underbrace{\frac{(1-\phi)\theta}{Nk_0}}_p \cdot k^2+\underbrace{\frac{(1-\phi)\theta}{Nk_0}(1-N)}_q\cdot k].\label{eq:D12}
\end{align}

Next, the sums in Eq.~(\ref{eq:D9}) can then be simplified to
\begin{align}
\sum\limits_{k=1}^{n}\prod_{j=1}^k\gamma_j=&n-\delta\sum\limits_{k=1}^n(pk^2+qk)\nonumber\\
=&n-\delta\left[p\cdot\frac{n(n+1)(2n+1)}{6}+q\cdot\frac{n(n+1)}{2}\right]\nonumber\\
=&n-\delta\left[\frac{(1-\phi)\theta}{6Nk_0}n(n+1)(2n+4-3N)\right].\label{eq:D13}
\end{align}

When substituting $n=j-1$ and $n=N-1$ into Eq.~(\ref{eq:D13}), we obtain the numerator and denominator of the right hand of Eq.~(\ref{eq:D9}). The fixation probability $\Phi_j$ is given by
\begin{align}
\Phi_j=&\frac{j+\delta\cdot\frac{(1-\phi)\theta}{6Nk_0}\cdot j(j-1)(3N-2j-2)}{N+\delta\cdot\frac{(1-\phi)\theta}{6k_0}(N-1)(N-2)}\nonumber\\
\approx&\!\frac{j}{N}\!+\!\frac{(1\!-\!\phi)j}{6N^2k_0}\theta\underbrace{\left[(j\!-\!1)(3N\!-\!2j\!-\!2)\!-\!(N\!-\!1)(N\!-\!2)\right]}_\Gamma\delta.\label{eq:D14}
\end{align}

\bibliography{aipsamp}

\end{document}